%% file: main.tex
\pdfoutput=1 
\documentclass[12pt]{article}

\usepackage[english]{babel}

\usepackage{xcolor}
\usepackage{pslatex}
\usepackage{amsmath}
\usepackage{amssymb}
\usepackage{bbm}
\usepackage{euscript}
\usepackage{epsfig}
\usepackage{graphicx}
\usepackage{setspace}
\usepackage{cite}

\newcommand{\mincas}{{\sf MINCAS}}
\newcommand{\katie}{{\sf KATIE}}

\newcommand{\alert}{\bf \color{red}}

\usepackage{listings}

\usepackage{epsfig}
\usepackage{amsmath}
\usepackage{amssymb}
\usepackage{graphicx}
\usepackage{cite}
\def\independenT#1#2{\mathrel{\setbox0\hbox{$#1#2$}%
 \copy0\kern-\wd0\mkern4mu\box0}} 
\def\desepsf(#1 width #2){\epsfxsize=#2 \epsfbox{#1}}

\def\be{\begin{equation}}
\def\ee{\end{equation}}
\def\bea{\begin{eqnarray}}
\def\eea{\end{eqnarray}}
\def\bsp{\begin{split}}
\def\esp{\end{split}}
\input{definitions.tex}

\definecolor{kkcolor}{rgb}{1,0,0}

\newcommand\kkout{\marginpar{\color{kkcolor}$\clubsuit$}\bgroup\markoverwith{\color{kkcolor}{\rule[04ex]{2pt}{0.8pt}}}\ULon}

\begin{document}
\begin{titlepage}

\begin{center}
\end{center}
\begin{flushright}
\bf IFJPAN-IV-2019-15 \\
    November 2019
\end{flushright}

\vspace{5mm}
\begin{center}
    {\Large\bf 
Jet quenching and effects of non-Gaussian transverse-momentum
broadening on di-jet observables$^{\star}$} 
\end{center}

\vskip 5mm
\begin{center}
{\large 
A. van Hameren$^a$,
K.\ Kutak$^a$,
        W.\ P\l{}aczek$^b$,
     M. Rohrmoser$^a$ and  
     K. Tywoniuk$^{c,\dagger}$
        }
\\
\vskip 2mm
{\em $^a$Institute of Nuclear Physics, Polish Academy of Sciences,\\
  ul.\ Radzikowskiego 152, 31-342 Krak\'ow, Poland}
\\
\vspace{1mm}
{\em $^b$Institute of Applied Computer Science, Jagiellonian University,\\
ul.\ \L{}ojasiewicza 11, 30-348 Krak\'ow, Poland}
\\
\vspace{1mm}
{\em $^c$ Department of Physics and Technology, University of Bergen, 5007 Bergen, Norway}
\end{center}
 
\vspace{20mm}
\begin{abstract}
\noindent
We study, at a qualitative level, production of jet pairs in ultrarelativistic nuclear collisions.
We propose  a new framework for combining $k_T$-factorisation and a formalism for in-medium propagation of jet particles that takes into account stochastic transverse forces as well as medium-induced radiation. This approach allows to address di-jet observables accounting for exact kinematics of the initial state. Using our framework, we provide a description of $R_{AA}$ data and study azimuthal decorrelations of the produced di-jets.   
In particular, we find that the resulting di-jet observables feature behaviour deviating from that of jet-pairs which undergo transverse-momentum broadening following the Gaussian distribution. We interpret this behaviour as a consequence of dynamics encoded in the Blaizot--Dominguez--Mehtar-Tani--Iancu equation.
\end{abstract}

\vspace{40mm}
\footnoterule
\noindent
{\footnotesize
$^{\star}$This work is partly supported by
 the Polish National Science Centre grant no. DEC-2017/27/B/ST2/01985.
}\\
\noindent
{\footnotesize
$^{\dagger}$KT is supported by a Starting Grant from Trond Mohn Foundation (BFS2018REK01) and the University of Bergen.
}

\end{titlepage}

\section{Introduction}

A prominent feature of high-energy hadronic collisions is the abundant jet production which is a manifestation of the underlying QCD dynamics. Jets are loosely defined as collimated sprays of particles that act as proxies for the properties of highly virtual partons, quarks and gluons, that participate in the hard scattering. Events where two jets approximately balance their momenta give an additional handle on probing how  initial-state processes and their associated parton distribution functions affect the properties of the final-state jets. It is important to point out that such vacuum effects lead to an appreciable azimuthal decorrelation as well as an imbalance of the transverse momentum of the leading and sub-leading, recoiling, jets.

Jet production in ultrarelativistic nucleus--nucleus collisions has a prominent role in probing the properties of the hot and dense nuclear matter formed in these events \cite{dEnterria:2009xfs, Mehtar-Tani:2013pia, Blaizot:2015lma}. 
This leads to the suppression, or 
quenching, of high-$p_T$ hadron and inclusive jet spectra observed both at $\sqrt{s_{NN}} = 200\,$GeV collisions at RHIC and $\sqrt{s_{NN}} = 5\,$TeV collisions at LHC \cite{Aad:2010bu,Chatrchyan:2011sx} (for a review see \cite{Qin:2015srf}).  It was early established experimentally that the modifications arose due to final-state interactions. This lead to the theoretical development by Baier--Dokshitzer--Mueller--Peigne--Schiff and Zakharov for the in-medium stimulated (bremsstrahlung) emissions that typically is referred to as the BDMPS-Z formalism \cite{Baier:1996kr,Baier:1996sk,Zakharov:1996fv,Zakharov:1997uu}\footnote{For an equivalent treatment within the thermal field theory see Ref.~\cite{Arnold:2002ja}.}.
Such emissions are responsible for transporting energy rapidly away from the jet axis to large angles \cite{Blaizot:2013hx,Kurkela:2014tla}. For high-$p_T$ jets, the total energy loss depends on the fragmentation properties of the jet \cite{Mehtar-Tani:2017web,Caucal:2018dla} (for the Monte Carlo implementation of this results see Ref.~\cite{Caucal:2019uvr}).

In the BDMPS-Z formalism, the medium affects the jet propagation and radiation via transverse momentum exchanges. Typical interactions are described by a diffusion constant $\hat q$. In a hot quark--gluon plasma (QGP) it is sensitive to its collective energy density. However, transverse momentum exchange between the jet and the medium are also expected. These are manifestations of the quasi-particles of the hot and dense matter and affect both the spectrum of radiated gluons \cite{CaronHuot:2010bp,Feal:2018sml,Feal:2018bru,Feal:2019xfl,Ke:2018jem,Mehtar-Tani:2019tvy,Mehtar-Tani:2019ygg} as well as the distribution of particles in transverse momentum space \cite{DEramo:2010wup,DEramo:2012uzl}. An important question is whether jet observables are sensitive to such interactions, especially those that are sensitive to recoils.
For example, the final-state interactions would lead to the gradual decorrelation of jets that originally were created from a vacuum $2\to 2$ matrix element \cite{Deak:2017dgs}, see also \cite{Mueller:2016gko,Jia:2019qbl,Ringer:2019rfk}.
In addition, one needs to account for  the initial state, i.e.\ evolution of the system that leads to hard scattering. It is therefore pertinent to further investigate how the details of the initial state affect the properties of the final state. In particular, how the
transverse-momentum dependence of the partons initiating the hard collision affect the 
azimuthal-angle decorrelations of the final-state jets, or dependence of $R_{AA}$ on the final-state transverse momentum. In approaches that account also for jet quenching the early stage of heavy-ion (HI) collisions  is usually described by the collinear factorisation where parton densities obey the DGLAP evolution equation and the initial-state partons are on mass-shell. Consequently, the final-state partons are essentially  produced back to back. To account for a non-vanishing transverse-momentum imbalance of the final-state partons, one often uses, on top of medium effects, initial-state parton showers via application of Monte Carlo generators, see e.g.\  \cite{Salgado:2003gb,Zapp:2008gi,Armesto:2009fj,Schenke:2009gb,Lokhtin:2011qq}. 

In this paper we propose an approach based on the combination of $k_T$-factorisation\cite{Catani:1990eg}, accounting for the longitudinal and transverse-momentum dependence of matrix elements and parton densities of the initial-state partons\footnote{ For recent phenomenological applications see \cite{Deak:2018obv,Bury:2017jxo,Kutak:2016ukc} and references therein.}, with the evolution in terms of the rate equation based on the final-state jet--plasma interaction.
Such an approach allows already at the lowest order, and without application of the initial-state shower (at least up to moderate values of transverse momenta), for a detailed study of the influence of kinematics of the initial state on the properties of the final-state system. We limit ourselves to study observables produced in mid-forward rapidity region and, for now, do not account for the initial-state saturation effects \cite{Gribov:1984tu,Gelis:2010nm,Kovchegov:2012mbw}.
Our focus here is to see to what extent the non-Gaussian spectrum of minijets as obtained in Ref.~\cite{Kutak:2018dim} is visible in the final-state observables. The study is rather of theoretical nature since we only account for gluonic jets. The equations for quarks have not been formulated yet. However, we believe that in the appropriate rapidity range, i.e.\ mid--forward region, the observables we are studying are mostly sensitive to gluons but still not to the extent to account for saturation effects  \cite{Gribov:1984tu,Gelis:2010nm,Kovchegov:2012mbw, vanHameren:2014ala}.
In order to describe the propagation of partons produced in hard collisions inside a hot quark--gluon plasma, we have applied the recently developed Monte Carlo generator \mincas~\cite{Kutak:2018dim} that solves the rate equation describing the rescattering and radiation of a hard parton in a dense QCD medium
\cite{Blaizot:2014rla}; for a similar approach see \cite{Feal:2018sml,Feal:2018bru}.

The paper is organised as follows. In Section~2 we present the theoretical framework of our study. In Section~3 numerical results of our Monte Carlo simulations are presented and discussed.
Section~4 concludes this work.
Finally, in Appendix~A we briefly describe the algorithm used in the numerical simulations.

\section{Theoretical framework}
\begin{figure}[t!]
\begin{center}
\includegraphics[width=0.45\textwidth]{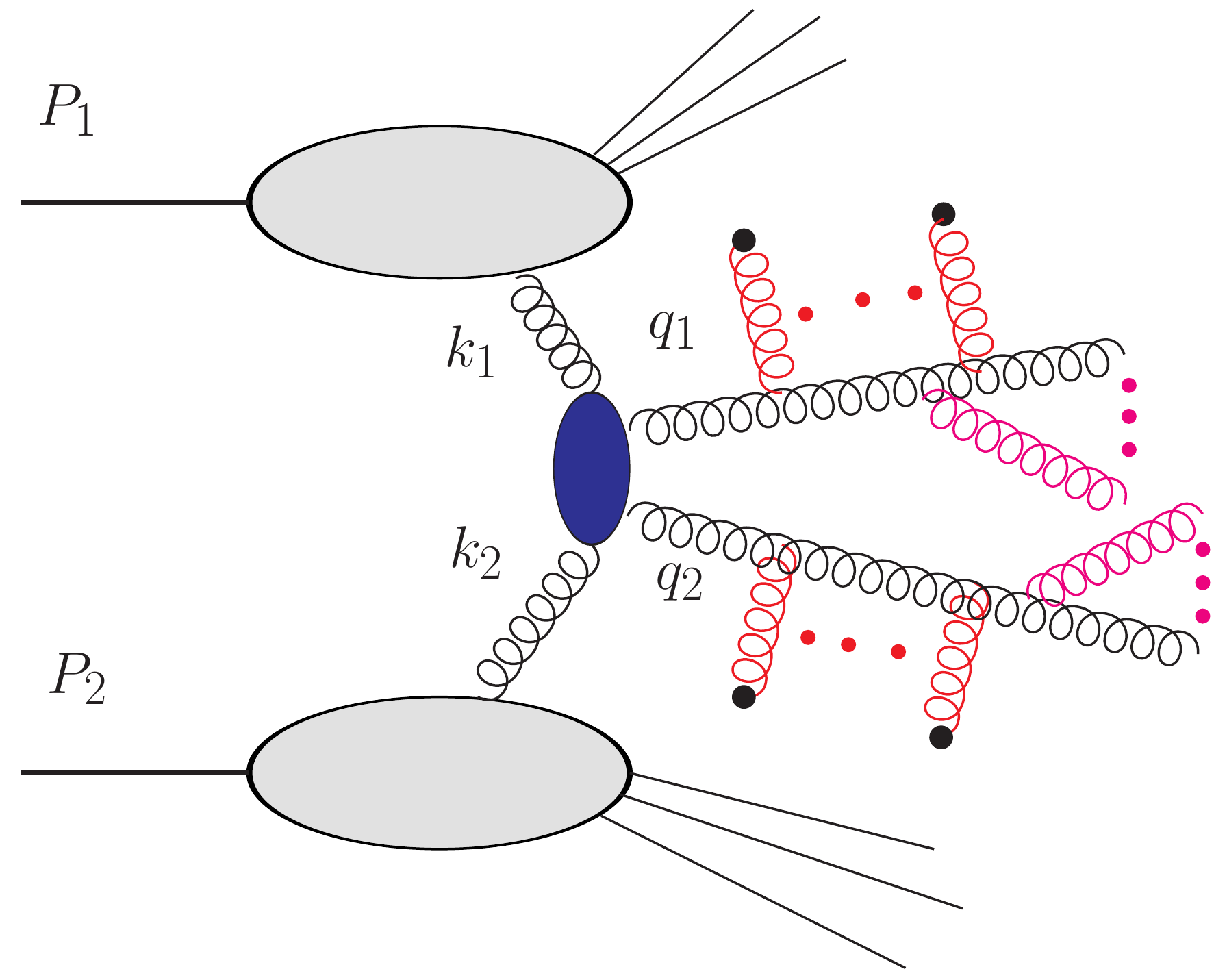}
\caption{
Gluon-jet production via two scattering gluons in hard nuclear collisions: colliding nuclei (horizontal ellipse) with the momenta $P_1$ and $P_2$ yield incoming gluons (with the momenta $k_1$ and $k_2$) which interact in a hard scattering process (vertical ellipse) and yield two gluons (with the momenta $q_1$ and $q_2$), which are subject to in-medium scattering (gluon interaction with $\bullet$), while simultaneously fragmenting into jets denoted by the purple colour.
\label{fig. hardprocess}}
\end{center}
\end{figure}

%
We factorise the production of a pair of gluon jets in nuclear collisions into the production of a pair of gluons and their subsequent in-medium evolution into gluon jets. 
The first step is described as the production of two gluons $G_1$ and $G_2$ via the hard collision of two gluons $G_A$ and $G_B$ which stem from two nuclei $N_A$ and $N_B$. 
We propose to describe this part of the process using the $k_T$-factorisation \cite{Catani:1989yc,Catani:1990eg}. 
The advantage of this approach is that it allows us to have access to the full phase space already at the LO accuracy.
The second step is given by the processes of scattering and medium-induced radiation that lead to the fragmentation of $G_1$ ($G_2$) into a jet $j_1$ ($j_2$).
Thus, the total process can be summarised as 
\begin{equation}
    N_A+N_B\rightarrow G_A + G_B + X\rightarrow G_1 + G_2 + X \rightarrow  j_1+j_2+X\,,
\end{equation}
where $X$ is the production of additional particles which are not used in our descriptions of (di)jet observables. 
The entire process is depicted schematically in Fig.~\ref{fig. hardprocess}.
This sections proceeds by first detailing the hard process $N_1+N_2\rightarrow G_1 + G_2 + X$, then the in-medium propagation $G_i\rightarrow j_i$ ($i=1,\,2$), followed by a short description of the modelling of the medium.

Jet production at mid-rapidity in proton-proton and heavy-ion collisions is typically computed within collinear factorisation. In contrast to most of the existing literature on jets, with the exception \cite{Deak:2017dgs}, in this work we consider (relatively) forward jet production within high-energy, or $k_T$, factorisation. This allows to introduce transverse-momentum dependent (also called unintegrated) parton distributions that resum effects of initial-state showering and the elementary partonic cross section between off-shell partons.

In this approach, the hard coefficient functions are calculated with space-like initial-state partons. The gauge invariance is guaranteed by adding the so-called induced terms \cite{Lipatov:1995pn,Lipatov:1996ts}, which equivalently can be viewed as embedding an off-shell amplitude in an on-shell one of higher order and taking the eikonal limit to decouple auxiliary on-shell lines while preserving terms which guarantee the gauge invariance \cite{vanHameren:2012if,vanHameren:2012uj}. 

This factorisation is therefore well suited to separate contributions to di-jet momentum imbalance and angular decorrelation from the initial-state and from final-state radiation and interactions. This is the main motivation for our current exploratory study. We would like to mention that this type of factorisation allows to be consistently augmented with a realistic parton shower~\cite{Bury:2017jxo}.

\subsection{$k_T$-factorisation of hard processes}
In the $k_T$-factorisation the initial state process reads
\begin{equation}
N_A(P_1) + N_B(P_2)\rightarrow G_A (k_1)+G_B(k_2)+X \rightarrow G_1(q_1)+G_2(q_2) +X\,,
\end{equation}
where the momenta $k_1$ and $k_2$ have components transverse to that of the incoming nuclei, an essential property for the description of the di-jet observables presented below, 
\begin{align}
k_1=x_1\,P_1+k_{1T},&& k_2=x_2\,P_2+k_{2T}\,. 
\end{align}
The momentum fractions $x_i$ and transverse momenta $k_{iT}$ follow the transverse-momentum distributions for gluons in both of the colliding nuclei given at a certain factorisation scale $\mu_F$.

Thus, the$k_T$-factorisation formula for the
parton-level differential cross section $\sigma_{pp}$ at the tree-level for the $g g$-pair production reads
\begin{eqnarray}\label{LO_kt-factorisation} 
\frac{d \sigma_{pp}}{d y_1 d y_2 d^2q_{1T} d^2q_{2T}} &=&
\int \frac{d^2 k_{1T}}{\pi} \frac{d^2 k_{2T}}{\pi}
\frac{1}{16 \pi^2 (x_1 x_2 s)^2} \; \overline{ | {\cal M}^{\mathrm{off-shell}}_{g^* g^* \to g g} |^2}
 \\  
&& \times  \; \delta^{2} \left( \vec{k}_{1T} + \vec{k}_{2T} 
                 - \vec{q}_{1T} - \vec{q}_{2T} \right) \;
{\cal F}_g(x_1,k_{1T}^2,\mu_{F}^2) \; {\cal F}_g(x_2,k_{2T}^2,\mu_{F}^2) \; \nonumber ,   
\end{eqnarray}
where ${\cal M}^{\mathrm{off-shell}}_{g^* g^* \to g \bar g}$ is the off-shell matrix element for the hard subprocess and ${\cal F}(x_i,k_{iT}^2,\mu_{F}^2)$ is the unintegrated gluon density (later on called the transverse-momentum-dependent gluon density) which, depending on an approximation used, obeys the Balitsky--Fadin--Kuraev--Lipatov (BFKL) \cite{Fadin:1975cb,Balitsky:1978ic}, Catani--Ciafaloni--Fiorani--Marchesini (CCFM) \cite{Catani:1989yc} or Balit\-sky--Kovchegov (BK) \cite{Balitsky:1995ub,Kovchegov:1999yj} equation or is given by some model, like that of Golec-Biernat--Wusthoff (GBW) \cite{GolecBiernat:1998js} or Kimber--Martin--Ryskin--Watt (KMRW) \cite{Martin:2009ii}. For the purpose of this work where we address rather moderate  values of longitudinal momenta fractions $x$ of partons in incoming hadrons, we use the Martin--Ryskin--Watt (MRW) gluon densities obtained from the {\sf CT10NLO} parton distribution functions (PDFs) via the application of the Sudakov form-factor\footnote{In the current study, to have a clear picture of broadening due to the non-Gaussian effects, we use just proton TMDs 
both for the proton--proton and nucleus--nucleus (A--A) collisions, but in the future we  plan to use nTMDs for the A--A case \cite{Blanco:2019qbm}.}.
The momentum fractions $x_i$, the rapidities $y_i$ and the transverse momenta $q_{iT}$ of the outgoing particles are related to one another as
\begin{equation}
x_1 = \frac{q_{1T}}{\sqrt{s}}\exp( y_1) 
     + \frac{q_{2T}}{\sqrt{s}}\exp( y_2)\,, \;\;\;\;\;\;
x_2 = \frac{q_{1T}}{\sqrt{s}}\exp(-y_1) 
     + \frac{q_{2T}}{\sqrt{s}}\exp(-y_2)\,. \nonumber
\end{equation}
For the numerical calculation of the hard-process cross sections we rely on the \katie\ framework \cite{vanHameren:2016kkz} which allows for evaluation of matix elements and cross sections with off-shell initial-state partons \cite{vanHameren:2012if,vanHameren:2012uj}.

\subsection{In-medium evolution}
In this subsection we describe the processes of the jet evolution in the medium,
\begin{equation}
G_i(q_i)\rightarrow j_i(p_i)\qquad i=1,2\,.
\end{equation}
The momentum of the jet is modified by its interactions with the medium. We have 
$\p=\ll+\q$, where the $\ll$ is the change of the jet transverse momentum and  $p^+=\tilde{x}q^+$ is the change of its longitudinal momentum due to momentum exchange with the medium and from the medium-induced branchings\footnote{Throughout this article, we use the following notation: The index `T' denotes the momentum components transverse to the beam axis of nuclear collisions, while symbols in bold-face represent the momentum components transverse to the jet-axis of one of the produced jets. The exception to this convention is the broadening of momenta transverse to the jet-axis which we call the ``$k_T$-broadening'', in order to be in agreement with many papers written on this subject.}.
To account for quenching, the cross section $\sigma_{pp}$ for the production of the vacuum jets should be convoluted with the fragmentation function $D(\tilde{x},\ll,\tau)$ for the jets in the medium.
Thus, for the inclusive production of the individual jets, the cross-section $d\sigma_{AA}/d\Omega_p$ for emission of jets into the phase-space element $d\Omega_p=dp_+d\p$ can be proposed
\begin{align}
\label{eq:inclusive-cross-section-1}
\frac{\dd \sigma_{AA}}{\dd \Omega_p} &= \int \dd \Omega_q \int \dd^2 \ll \int_0^1 \frac{\dd \tilde{x}}{\tilde{x}} \, \delta(p^+ - \tilde{x} q^+) \delta^{(2)}(\p-\ll-\q)\, D\left(\tilde{x},\ll,\tau(q^+)\right) \frac{\dd \sigma_{pp}}{\dd \Omega_q}  \nn
&= \int \dd^2\q \int_0^1 \frac{\dd \tilde{x}}{\tilde{x}^2} D\left(\tilde{x},\p-\q,\tau(p^+/\tilde{x})\right) \left.\frac{\dd \sigma_{pp}}{\dd q^+ \dd^2 \q} \right\vert_{q^+ = p^+/\tilde{x}},
\end{align}
where $\dd \Omega_q = \dd q^+ \dd^2\q$, $\tau(q^+) = \bar \alpha \sqrt{\hat q/q^+}L$, and
\beq
D(\tilde{x},\ll,\tau) \equiv \tilde{x} \frac{\dd N}{\dd \tilde{x}\, \dd^2\ll} \,,
\eeq
is the distribution of gluons with momentum fraction $\tilde x$ and transverse momentum ${\bs l}$ (relative to the jet axis) after passing through a medium with length $L^+$ (as encoded in the re-scaled evolution variable $\tau$).
Similarly for the production of di-jets, the differential cross-section for the emission of two jets into the phase-space elements $d\Omega_{p_1}$ and $d\Omega_{p_2}$ can be written as
\begin{align}
\label{eq:inclusive-cross-section-2}
\frac{\dd \sigma_{AA}}{\dd \Omega_{p_1}\Omega_{p_2}} = & \int \dd \Omega_{q_1} \dd \Omega_{q_2}\int \dd^2 \ll_1  \int \dd^2 \ll_2
\int_0^1 \frac{\dd \tilde{x}_1}{\tilde{x}_1} \, \delta(p^+_1 - \tilde{x}_1 q^+_1)\int_0^1 \frac{\dd \tilde{x}_2}{\tilde{x}_2} \, \delta(p^+_2 - \tilde{x}_2 q^+_2)\nonumber\\
&\times \delta^{(2)}(\p_1-\ll_1-\q_1)\,\delta^{(2)}(\p_2-\ll_2-\q_2)\,D(\tilde{x}_1,\ll_1,\tau(q^+_1)) D(\tilde{x}_2,\ll_2,\tau(q^+_2))\frac{\dd \sigma_{pp}}{\dd \Omega_{q_1}\dd \Omega_{q_2}} \, \nn
= &\int \dd^2\q_1  \int \dd^2\q_2\int_0^1 \frac{\dd \tilde{x}_1}{\tilde{x}^2_1} \int_0^1 \frac{\dd \tilde{x}_2}{\tilde{x}^2_2} D(\tilde{x}_1,\p_1-\q_1,\tau(p^+_1/\tilde{x}_1)) D(\tilde{x}_2,\p_2-\q_2,\tau(p^+_2/\tilde{x}_2))\\\nonumber
&\times \left.\frac{\dd \sigma_{pp}}{\dd q^+_1\dd q^+_2\dd^2 \q_1\dd^2 \q_2} \right\vert_{q^+_1 = p^+_1/\tilde{x}_1,\,q^+_2 = p^+_2/\tilde{x}_2}\,,
\end{align}
where it is assumed implicitly that the fragmentation processes of the jet~$1$ and the jet~$2$ factorise from the hard scattering process as well as from each other. 

The evolution equation for the gluon transverse-momentum-dependent
distribution $D(\tilde{x},\mathbf{l},t)$ in the dense medium, 
obtained under the assumption that the momentum transfer in the kernel is small, reads
\cite{Blaizot:2014rla}
\begin{equation}
\begin{aligned}
\frac{\partial}{\partial t} D(\tilde{x},\mathbf{l},t) = & \: \frac{1}{t^*} \int_0^1 dz\, {\cal K}(z) \left[\frac{1}{z^2}\sqrt{\frac{z}{\tilde{x}}}\, D\left(\frac{\tilde{x}}{z},\frac{\mathbf{l}}{z},t\right)\theta(z-\tilde{x}) 
- \frac{z}{\sqrt{\tilde{x}}}\, D(\tilde{x},\mathbf{l},t) \right] \\
+& \int \frac{d^2\mathbf{q}}{(2\pi)^2} \,C(\mathbf{q})\, D(\tilde{x},\mathbf{l}-\mathbf{q},t),
\end{aligned}
\label{eq:ktee1}
\end{equation}
where 
\begin{equation}
{\cal K}(z) = \frac{\left[f(z)\right]^{5/2}}{\left[z(1-z)\right]^{3/2}}, 
\quad   f(z) = 1 - z + z^2, 
\qquad  0 \leq z \leq 1, 
\label{eq:kernel1}
\end{equation}
is the $z$-kernel function, and
\begin{equation}
 \frac{1}{t^*}  = \frac{\bar{\alpha}}{\tau_{\rm br}(E)} = \bar{\alpha}\sqrt{\frac{\hat{q}}{E}}, 
\qquad \bar{\alpha} = \frac{\alpha_s N_c}{\pi},
\label{eq:tstar}
\end{equation}
where $t^*$ is the stopping time, i.e.\ the time at which the energy of an incoming parton has been radiated off in form of soft gluons, 
$E$ is the energy of the incoming parton, 
$z$ -- its longitudinal momentum fraction, 
$\hat{q}$ -- the quenching parameter,
$\alpha_s$ -- the QCD coupling constant
and $N_c$ -- the number of colours.
The kernel $\mathcal{K}(z)$ accounts for soft gluon emissions that are the dominant contribution to jet energy loss. On the other hand, the collision kernel $C(\mathbf{q})$ is given by
\begin{equation}
C(\mathbf{q}) = w(\mathbf{q}) - \delta(\mathbf{q}) \int d^2\mathbf{q'}\,w(\mathbf{q'})\,,
\label{eq:Cq}
\end{equation}
and includes perturbative, i.e. $\sim 1/{\bs q}^4$, re-scattering with the medium.
Here we consider a situation where the quark--gluon plasma equilibrates 
and the transverse-momentum distribution of medium particles assumes the form \cite{Aurenche:2002pd}
\begin{equation}
 w(\mathbf{q}) = \frac{16\pi^2\alpha_s^2N_cn}{\mathbf{q}^2(\mathbf{q}^2+m_D^2)}\,,
\label{eq:wq2}
\end{equation}
where $m_D$ is the Debye mass of the medium quasi-particles.
In the following we consider the expression of Eq.~(\ref{eq:wq2}) inside the collision kernel $C(\q)$.

The above integral equations can be formally solved by iteration. Denoting $\tau=t/t^*$, we get \cite{Kutak:2018dim}

\begin{equation}
\begin{aligned}
 D(\tilde{x},\mathbf{l},\tau)  =\:& \int_0^1 d\tilde{x}_0 \,\int d^2 \mathbf{l}_0 \, D(\tilde{x}_0,\mathbf{l}_0,\tau_0) \bigg\{ e^{-\Psi(\tilde{x}_0)(\tau - \tau_0)} \delta(\tilde{x}-\tilde{x}_0)\,\delta(\mathbf{l}-\mathbf{l}_0)\\
 +\:& \sum_{n=1}^{\infty}\prod_{i=1}^n \left[ \int_{\tau_{i-1}}^{\tau} d\tau_i \, \int_0^1 dz_i\, \int d^2\mathbf{q}_i\,
 {\cal G}(z_i,\mathbf{q}_i)\,e^{-\Psi(\tilde{x}_{i-1})(\tau_i - \tau_{i-1})} \right] \\
 &\times e^{-\Psi(\tilde{x}_n)(\tau - \tau_n)} \, \delta(\tilde{x}-\tilde{x}_n)\,\delta(\mathbf{l}-\mathbf{l}_n)\bigg\}\,,
\end{aligned}
\label{eq:ktItfin}
\end{equation}
where
\begin{eqnarray}
\Psi(\tilde{x}) &=&\Phi(\tilde{x}) + W\,,
\label{eq:Psi}\\
\Phi(\tilde{x}) &=& 
 \frac{1}{\sqrt{\tilde{x}}}  \int_0^{1-\epsilon} dz\, z {\cal K}(z),
\label{eq:Phi} \\
W &=& 
    t^* \int_{|\mathbf{q} |>q_{\mathrm{min}}} d^2 \mathbf{q}
    \,\frac{w(\mathbf{q})}{(2\pi)^2}, 
\label{eq:W} 
\end{eqnarray}
\begin{equation}
{\cal G}(z,\mathbf{q}) = \sqrt{\frac{z}{\tilde{x}}} \, z{\cal K}(z)\,\theta(1-\epsilon-z) \,\delta(\mathbf{q}) 
                                 +  t^* \, \frac{w(\mathbf{q})}{(2\pi)^2} \,\theta(|\mathbf{q} | - q_{\mathrm{min}}) \delta(1-z) \, ,
\label{eq:ktKer}
\end{equation}
and
\begin{equation}
\tilde{x}_n = z_n \tilde{x}_{n-1},\qquad 
\mathbf{l}_n = z_n \mathbf{l}_{n-1} + \mathbf{q}_n \,,
\label{eq:xnkn}
\end{equation}
with $\tilde{x}_0$ and $\mathbf{l}_0$ being some initial values of 
$\tilde{x}$ and $\mathbf{l}$ at the initial evolution time $\tau_0$, 
given by the distribution $D(\tilde{x}_0,\mathbf{l}_0,\tau_0)$.

Furthermore, after integration of Eq.~(\ref{eq:ktee1}) over the transverse
momentum $\mathbf{l}$ one obtains the evolution equation for
the gluon energy density \cite{Blaizot:2014rla}:
\begin{equation}
\begin{aligned}
\frac{\partial}{\partial t} D(\tilde{x},t) = & \: \frac{1}{t^*} \int_0^1 dz\, {\cal K}(z) \left[\sqrt{\frac{z}{\tilde{x}}}\, D\left(\frac{\tilde{x}}{z},t\right)\theta(z-\tilde{x}) 
- \frac{z}{\sqrt{\tilde{x}}}\, D(\tilde{x},t) \right] \,,
\end{aligned}
\label{eq:qee1}
\end{equation}
where $D(\tilde{x},t)\equiv \int \dd^2\ll\, D(\tilde{x},\ll,t)$. 
The iterative solution of this equation reads \cite{Kutak:2018dim}
\begin{equation}
\begin{aligned}
D(\tilde{x},\tau) & =  \int_0^1 d\tilde{x}_0 \, D(\tilde{x}_0,\tau_0) \bigg\{  e^{-\Phi(\tilde{x}_0)(\tau - \tau_0)}\, \delta(\tilde{x} - \tilde{x}_0) \\
    & + \sum_{n=1}^{\infty} \prod_{i=1}^n  \left[ \int_{\tau_{i-1}}^{\tau} d\tau_i  \int_0^1  dz_i\, \sqrt{\frac{z_i}{\tilde{x}_i}} \,z_i{\cal K}(z_i)\, \theta(1-\epsilon-z_i)\,e^{-\Phi(\tilde{x}_{i-1})(\tau_i - \tau_{i-1})} \right]
    \\
    & \times e^{-\Phi(\tilde{x}_n)(\tau - \tau_n)} \, \delta(\tilde{x} - \tilde{x}_n) \bigg\}\,.
\end{aligned}
\label{eq:iterx}
\end{equation}
Both Eqs.~(\ref{eq:ktee1}) and (\ref{eq:qee1}) are solved numerically within the \mincas\ framework with the use of dedicated Markov Chain Monte Carlo (MCMC) algorithms \cite{Kutak:2018dim}.
In this article, we generally evolve the gluon di-jets following the $k_T$-dependent evolution equation~(\ref{eq:ktee1}) and compare the results to the di-jet evolution using Eq.~(\ref{eq:qee1}) combined with the Gaussian $k_T$-broadening, in order to study the effects of the non-Gaussian $k_T$-broadening.

In both cases, the initial condition for the evolution is a single particle, i.e. $D(\tilde x ,{\bs l}, \tau=0) = \delta(1-x) \delta({\bs l})$ (or $D(\tilde x, \tau = 0 ) = \delta(1-x)$). After the passage through the medium, medium-induced branching and broadening will generate a collection of particles described by the final distribution $D(\tilde x, {\bs l},\tau)$. The notion of the leading particle, which dominates the contribution to the inclusive cross sections, Eqs.~(\ref{eq:inclusive-cross-section-1}) and (\ref{eq:inclusive-cross-section-2}), can be recovered in the limit $\tilde x \approx 1$.

\subsection{Medium model}\label{ssec:mmodel}
Numerous approaches already exist that describe the evolution of the QGP-medium   with time, cf.\ Refs.~\cite{Florkowski:2010zz,Jaiswal:2016hex} and references therein.
However, rather than at a detailed quantitative description of phenomenology, our current study aims at qualitative understanding of the effects of $k_T$-broadening on di-jet evolution. 
For an isolated study of $k_T$-broadening  effects and for simplicity, we assume that the medium exists for some time $t_L$ with the constant temperature $T$ and is absent at later times. 
Thus, the medium depends only on two free parameters, while  the temperature dependencies of the medium properties that are necessary to describe the jet evolution following Eqs.~(\ref{eq:ktee1}) and~(\ref{eq:qee1}) can be obtained by phenomenological considerations.
The JET Collaboration~\cite{Burke:2013yra} has obtained the temperature dependence of the transport parameter $\hat{q}$ as 
\begin{equation}
    \hat{q}(T)=c_q T^3\,.
\end{equation}
The number of scattering centres can be estimated by assuming a medium consisting of fermions and bosons at the thermal equilibrium, i.e.\ by assuming the Fermi--Dirac/Bose--Einstein distributions for the densities of quarks, antiquarks and gluons, $n_q$, $n_{\bar{q}}$, and $n_g$, respectively. As can be shown, cf.\ e.g.\ Eq.~(3.14) in~\cite{Zapp:2008zz}, the Taylor-expansion in $T$ yields the number densities as the cubic power of $T$ at the lowest orders in $T$, so that one can write
\begin{equation}
    n(T)=n_q+n_{\bar{q}} +n_g =c_n T^3\,.
\end{equation}
For the Debye-mass $m_D$ we assume that $m_D\propto gT$, which is consistent with findings of the Hard-Thermal-Loop (HTL) approach. In particular, following~\cite{Laine:2016hma}, we use the relation
\begin{equation}
    m_D^2=\left(\frac{N_C}{3}+\frac{N_F}{6}\right) g^2T^2\,.
\end{equation}

In our setup, we have not yet included realistic geometry from the Glauber model for nucleus--nucleus collisions nor the effect of the expansion of the medium. This would add more fluctuations to both the path-length distributions as well as the temperature profile probed by the jets. Hence, all partons traverse the same length in the medium, $t_L = \mathrm{const.}$, and the same temperature profile. However, while a more realistic geometry certainly is needed to precisely extract medium parameters, these effects were found to be of secondary importance to understand di-jet acoplanarity in realistic Monte Carlo studies \cite{Milhano:2015mng}, and will therefore be revisited in the future.

\section{Numerical results}

\subsection{Jet-quenching and medium parameters}

As outlined in Subsection~\ref{ssec:mmodel}, the effective model for the medium and the jet--medium interactions depends on four parameters, which additionally yield three values that are currently used as parameters within \mincas. These parameters need to be tuned by comparison to experimental data.
Of the aforementioned five parameters, $c_q$ is fixed by phenomenological considerations described in~\cite{Burke:2013yra}. Also $c_n$ is fixed, as it is the first coefficient in the Taylor-series expansion of $n(T)$. 
Thus, there remain only the two free parameters $t_L$ and $T$. 
Assuming that a medium of a diameter of the order of $10\,$fm is created and that most jets are created in the centre of the colliding particles and pass the medium with a velocity close to the speed of light (i.e.\ ultrarelativistic jet particles), we set the parameter $t_L$ to $5\,$fm/c.
The temperature $T$  is then varied in order to reproduce experimental data on the jet-quenching.

With regard to the jets, we have made two essential approximations in order to be able to produce qualitative results for the observables:
\begin{enumerate}
    \item Only gluon jets have been considered, since currently evolution equations for quarks analogous to Eq.~(\ref{eq:ktee1}) are not known. To minimise possible errors due to the negligence of quark-jets, we consider only observables in a mid--forward rapidity region where gluon jets dominate.
    \item We identify the momentum of the gluon-jet with that of its leading particle. 
    To minimise the possible resulting errors, we consider only leading particles with $p_T$ above a certain threshold. 
\end{enumerate}

One of the most inclusive widely studied observable is the  nuclear modification ratio $R_{AA}$ as a function of $p_T$, 
\begin{equation}
    R_{AA}(p_T)=\frac{1}{\langle T_{AA}\rangle}\,\frac{dN_{AA}/dp_T}{d\sigma_{pp}/dp_T}\,,
\end{equation}
where $\langle T_{AA}\rangle$ is the average nuclear overlap function.
For the qualitative considerations of this work, nuclear effects of suppression or enhancement other than jet-quenching in the medium have been neglected and the nuclear modification factor is thus approximated as 

\begin{equation}
R_{AA}(p_T) \approx \frac{d\sigma_{AA}/dp_T}{d\sigma_{pp}/dp_T}\,, \end{equation}
where $d\sigma_{AA}/dp_T$ is normalised to the number of binary collisions of nucleons in the A--A collision.

\begin{figure}[t!]
    \centering
    \includegraphics[width=1.0\linewidth,clip=true, trim=25 0 45 0]{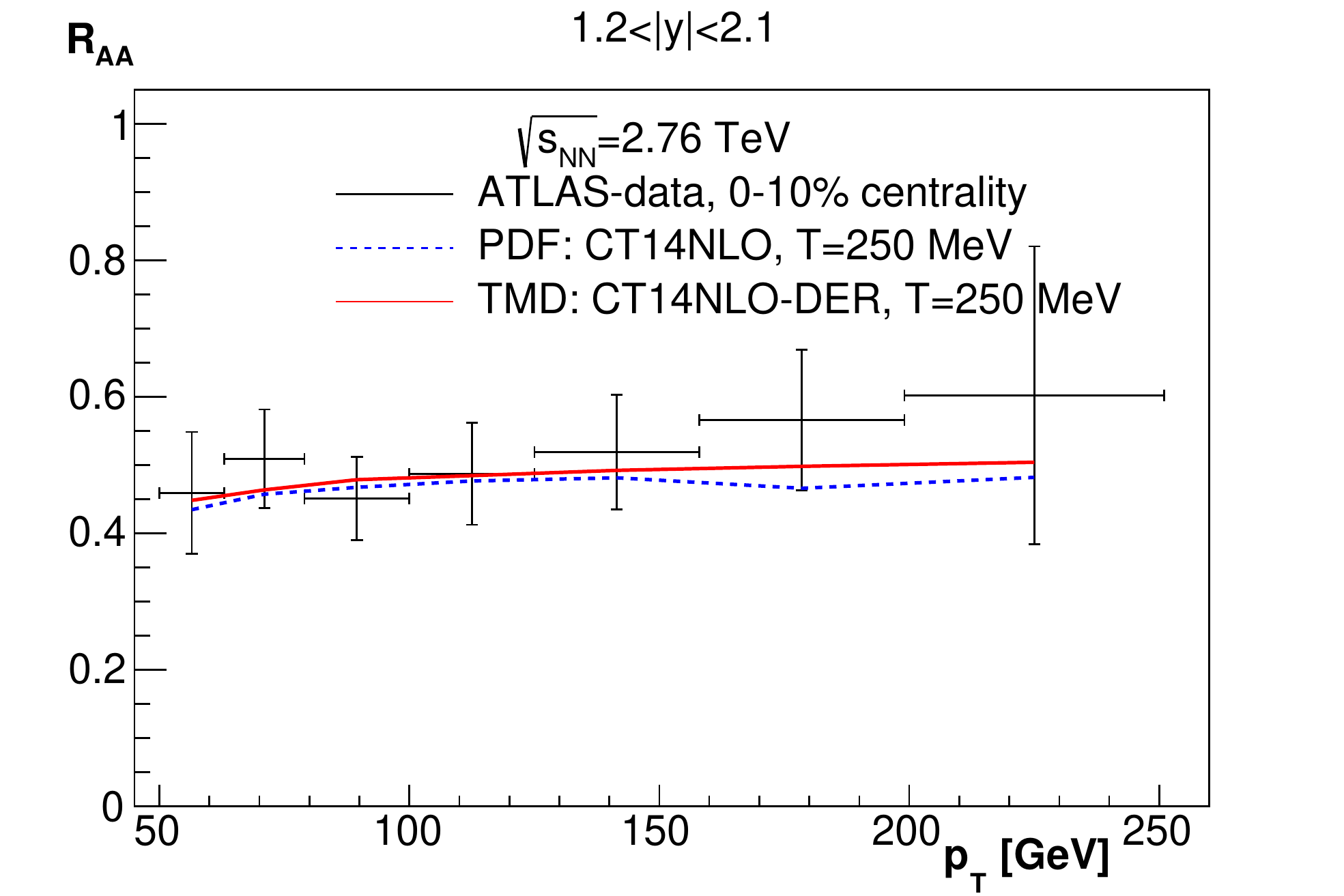}
    \caption{The nuclear modification factor $R_{AA}$ as a function of the jet transverse momentum $p_T$ for the Pb--Pb collisions at $\sqrt{s_{\rm NN}}=2.76\,$TeV in the mid-forward $1.2<|y|<2.1$ region, obtained with the initial-state collinear gluon density (PDFs) and transverse-momentum-dependent (TMD) gluon density as indicated in comparison to the LHC data taken from \cite{Aad:2014bxa}. }
    \label{fig:raa}
\end{figure}

In Fig.~\ref{fig:raa} we show the experimental data from ATLAS on $R_{AA}$ for jets in the Pb--Pb collisions at $\sqrt{s_{NN}}=2.76\,$TeV~\cite{Aad:2014bxa} in comparison with the results from jet simulations with our combination of the \katie\ and \mincas\ Monte Carlo generators. 
In order to reproduce the data, the value of $T$ is tuned to $250\,$MeV.
The results for $R_{AA}$ relying on the gluon TMDs and the gluon PDFs both exhibit a considerable suppression, corresponding to the values of $R_{AA}$ between $0.4$ and $0.6$, and, in general, show a similar behaviour.
However, it can be noted that at the same temperature scales, the results based on the gluon PDFs are at high $p_T$-scales slightly more suppressed than that for the case where gluon distributions depend on transverse momentum.
As can be seen, with the chosen value of the temperature our Monte Carlo predictions fit well the data points.

With tuning the temperature scales to the experimental results for $R_{AA}$, we have fixed all model parameters -- they are summarised in Table~\ref{tab:parameters}.

{\alert
\begin{table}[!htbp]
    \centering
    \begin{tabular}{||cl|cl|cl||}
    \hline\hline
    \multicolumn{2}{||c|}{fixed}&\multicolumn{2}{c|}{free}&\multicolumn{2}{c||}{resulting}\\\hline\hline
        $c_q$ & $3.7$ &$t_L$&$5\,~$fm/c & $\hat{q}$& $0.29\,~$GeV$^2$/fm\\\hline
         $c_n$& $5.228$&T& $0.25$~GeV &$ n$& $0.08\,~$GeV$^3$\\\hline
         && &&$m_D$& $0.61$~GeV\\
    \hline\hline
    \end{tabular}
    \caption{Parameters for the medium model: the parameters from theoretical/phenomenological considerations (left), the freely adjustable parameters (middle) and the resulting medium parameters used for \mincas\ (right).}
    \label{tab:parameters}
\end{table}
}

\subsection{Di-jet results}

\begin{figure}[!htbp]
    \centering
    \includegraphics[width=1.0\linewidth,clip=true, trim=25 5 45 0]{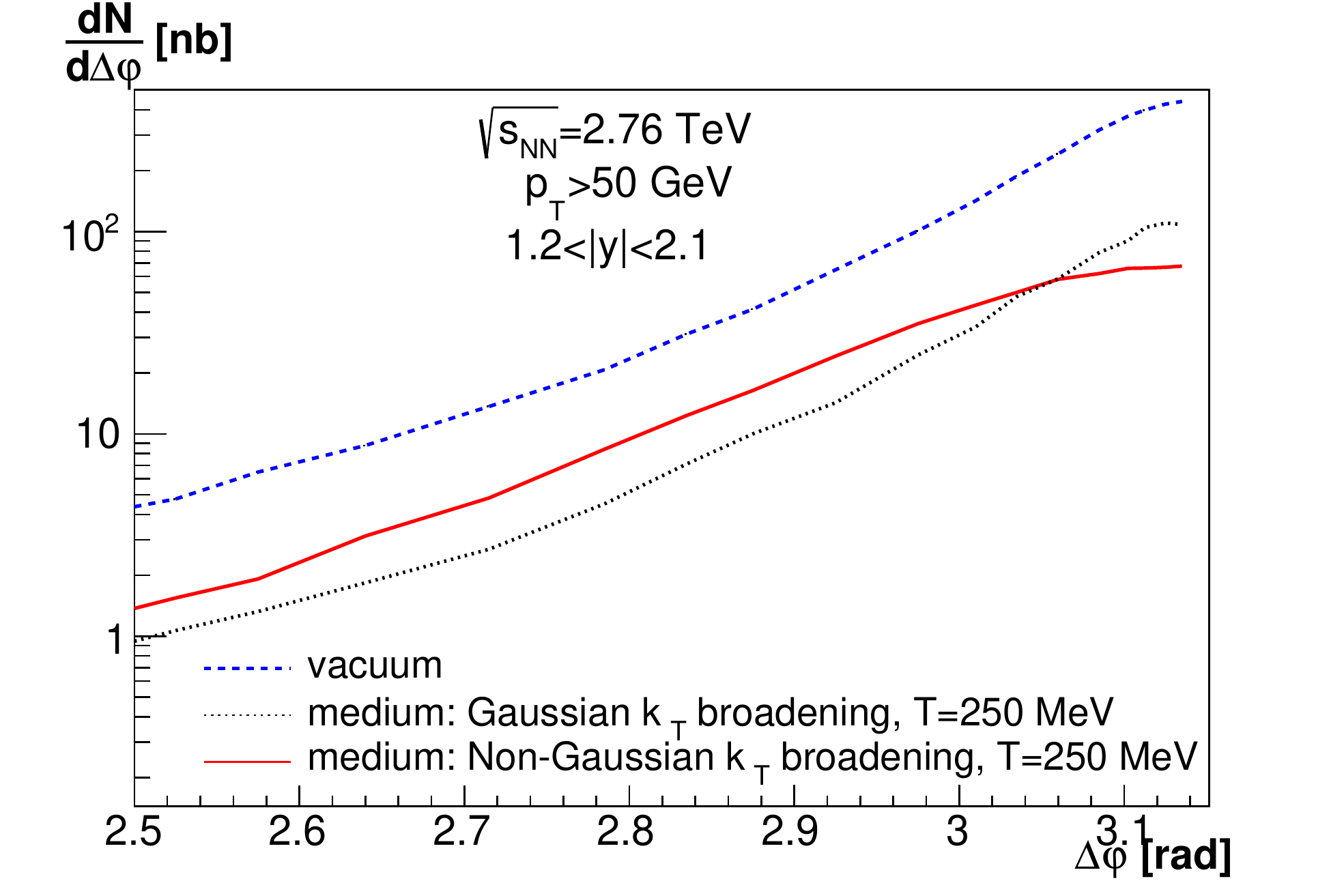}
    \caption{Azimuthal di-jet decorrelations for collisions at $\sqrt{s_{\rm NN}}=2.76\,$TeV between two jets, each with rapidity $1.2<|y|<2.1$, as indicated. The results are obtained for the proton--proton collisions (vacuum -- the dashed blue line) as well as for the Pb--Pb collisions with the $k_T$-broadening as in \mincas\  (Non-Gaussian -- the solid red line) and the Gaussian $k_T$-broadening (the dotted black line). 
}
    \label{fig:dphi}
\end{figure}

\begin{figure}[!htbp]
    \centering
    \includegraphics[width=1.0\linewidth,clip=true, trim=20 5 45 0]{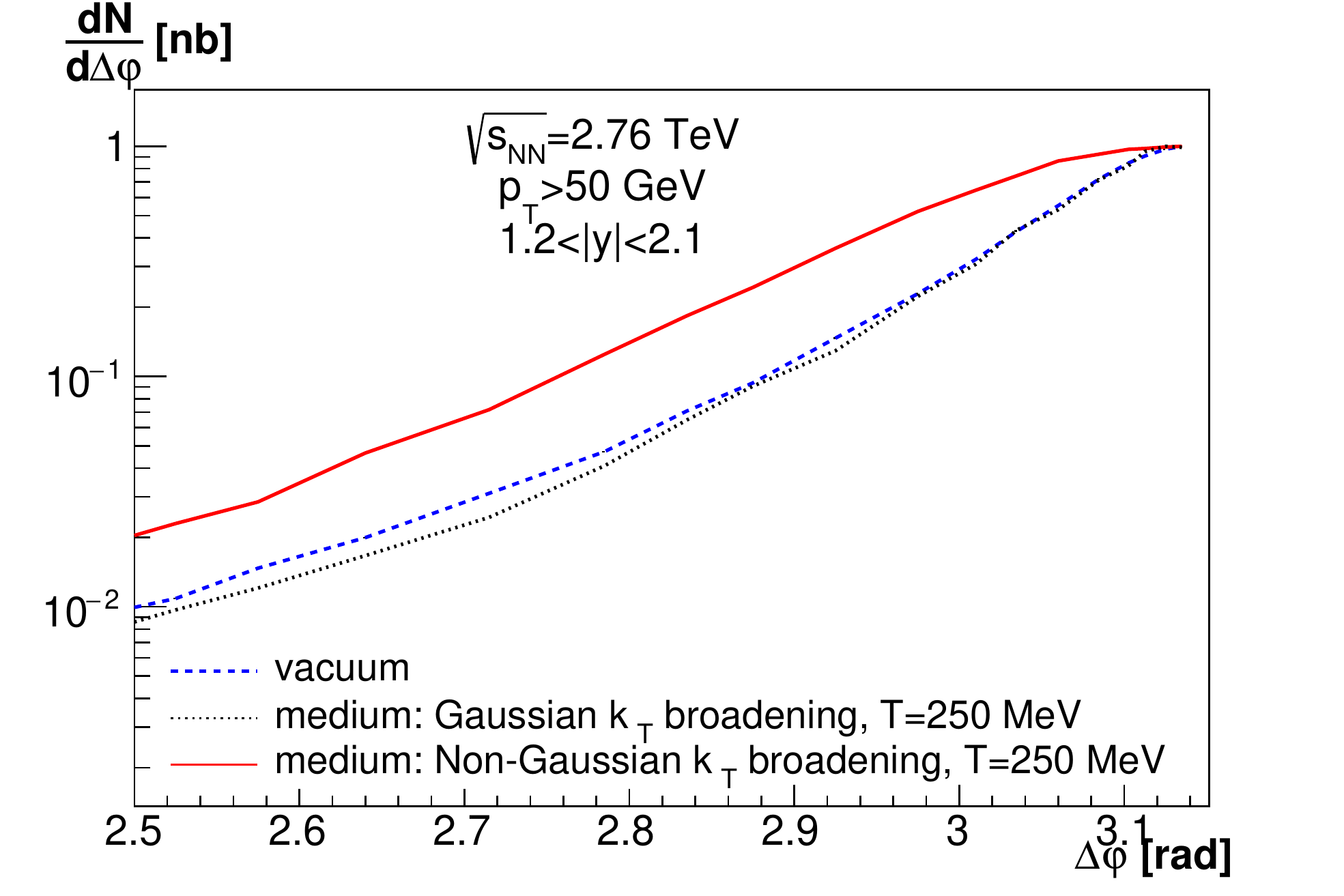}

    \caption{Same as in Fig.~\ref{fig:dphi}, but now normalised to the maximum of the distribution. 
}
    \label{fig:dphinorm}
\end{figure}


As has been already mentioned, the use of the $k_T$-factorisation  allows to study, already at the LO level, di-jet observables in the full phase space, i.e.\ to have access to regions away from the back-to-back configuration in the transverse plane. 
In our results, we compare the production of jet-pairs in hard collisions without and with further in-medium evolution. The former case, labelled as  the ``vacuum'' case, corresponds to, e.g. the di-jet production in the proton--proton collisions and was obtained numerically by the use of \katie\ alone. Thus, it already contains an asymmetry in the transverse momenta $k_T$ of the jets due to the use of the transverse-momentum-dependent gluon density instead of the gluon PDF.
The latter case, labelled as the ``medium'' case, may contain additional $k_T$-broadening effects of the jet-axes due to the jet--medium interactions.
We have obtained the results for the ``medium'' case by propagating within \mincas\ the gluons produced in the hard collisions by \katie, where the gluon-fragmentation functions follow Eqs.~(\ref{eq:ktee1}) and~(\ref{eq:qee1}).
To further investigate the in-medium $k_T$-broadening we have simulated two different cases:
\begin{enumerate}
    \item In the first case the jet in-medium fragmentation follows Eq.~(\ref{eq:ktee1}). Inside Eq.~(\ref{eq:ktee1}), $C(\q)$ yields the broadening of momenta transverse to the jet-axis. We call this case the ``non-Gaussian $k_T$-broadening''.
    \item In the second case the the loss of jet-momentum components along the jet-axis, $\vec{q}_i$ ($i=1,2$), follows Eq.~(\ref{eq:qee1}). Subsequently, the transverse momentum component $\ll_i\perp \vec{q}_i$ is selected (for each jet-momentum individually) from the Gaussian distribution. As it can be argued that the absolute value of the total transverse-momentum transfer is of the order of $\sqrt{\hat{q}t_L}$, $||\ll_i||$ is selected from the Gaussian distribution 
    \begin{equation}
        P(||\ll_i||)=\frac{1}{\sqrt{2\pi\hat{q}t_L}}\;
        \exp\left(-\frac{\ll_i^2}{2\hat{q}t_L}\right).
    \end{equation}
    The azimuthal angle of the outgoing momenta $p_i$ with regard to $q_i$ is selected randomly from a uniform distribution in the range from $0$ to $2\pi$.
    We label the resulting set of jets with the ``Gaussian $k_T$-broadening''.
\end{enumerate}

We study the azimuthal decorrelation $dN/d\Delta \Phi$ given as the number of jet-pairs, where $\Delta \Phi$ is the difference in azimuthal angles of the jet-axes. 
The results are shown in Fig.~\ref{fig:dphi}  for the di-jets where 
both jets are emitted with transverse momenta $p_T$ above a threshold of $50$~GeV and rapidities $y$ in the region $1.2<|y|<2.1$.
It can be seen that the production of the jet-pairs is clearly suppressed in the medium as compared to the production of the jets without the subsequent in-medium propagation.
While, compared to the vacuum, the $dN/d\Delta \Phi$ values in the medium are similarly suppressed, whether we assume the Gaussian or non-Gaussian $k_T$-broadening,  differences in the behaviour of both curves occur, which can be made more visible by normalising the curves for ${dN}/{d\Delta \Phi}$ to the values at their respective maximums $(dN/d\Delta\Phi)_{\rm max}$.
These results are shown in Fig.~\ref{fig:dphinorm}. 
The case with the non-Gaussian $k_T$-broadening exhibits a clear broadening in $\Delta \Phi$ as compared to the vacuum case, while the case with the Gaussian $k_T$-broadening mostly follows the behaviour of the vacuum case.

\section{Conclusions and outlook}

We have proposed a factorisation formula combining the $k_T$-factorisation and medium--jet interactions. Using this formula and its implementation in Monte Carlo generators, we have studied combined effects of transverse momenta of initial-state partons with transverse momenta in the final state generated due to medium--jet interactions. The latter features non-Gaussian behaviour due to interplay of radiation and in-medium scattering. The study has been performed using  Monte Carlo programs which combine both the hard-scattering process depending on transverse momenta of partons within nucleons and the in-medium evolution for jet-particles.
In order to allow for a reasonable jet in-medium evolution, we describe the medium with a simplified effective model that relies merely on two parameters: the time of in-medium jet-evolution $t_L$ and a constant medium temperature $T$.
This algorithm combines the previously developed frameworks of \katie~\cite{vanHameren:2016kkz} and \mincas~\cite{Kutak:2018dim}, and so far is restricted to the production of gluons only.

With the corresponding tuning of $T$, it has been possible to reproduce the experimental results for the jet nuclear modification factor $R_{AA}$ for the  Pb--Pb collisions at $\sqrt{s_{NN}}=2.76\,$TeV.
Using the algorithm in this calibration, we have been able to calculate (at a qualitative level, since quarks are missing in our study) the azimuthal di-jet decorrelations $dN/d\Delta \Phi$ in Pb--Pb collisions, where we have compared the results for the Gaussian $k_T$-broadening in the medium with that of the non-Gaussian $k_T$-broadening which follows from Eq.~(\ref{eq:ktee1}).  
Our framework is still too simplistic to really address the full quantitative description of experimental data, since we do not the include vacuum-like emissions in the medium nor the parton shower outside the medium \cite{Caucal:2018dla,Caucal:2019uvr}. However, the effect of the vacuum final-state parton  shower on $\Delta\phi$ was studied in Ref.~\cite{Bury:2016cue} and it was demonstrated that the  $\Delta\phi$ distributions obtained using the $k_T$-factorisation and {\sf PYTHIA} \cite{Sjostrand:2014zea} with the final-state parton shower (FSR) agree in shape, while they differ slightly in normalisation. When FSR was neglected the two curves were almost on top of each other. Therefore, we expect that this effect is universal for the vacuum and the medium, and does not influence the comparison of the shapes of both the results. 
We think that our main result for the azimuthal decorrelations, 
i.e.\ the non-Gaussian $k_T$-broadening,
is universal and leads to the considerable broadening of the shape of the $dN/d\Delta\phi$ distribution as compared to the Gaussian $k_T$-broadening, 
as well as to the case of the proton--proton collisions alone.

In the future, we plan to extend our framework to account for quarks which have been neglected in the current study.
This will allow us to apply it to phenomenology focused on testing the pattern of jet-quenching 
in jet--jet, jet--hadron \cite{Adam:2015doa,Adamczyk:2017yhe} and jet--electroweak-boson \cite{Sirunyan:2017jic,ATLAS:2019gif} final-state systems.

Furthermore, we plan to investigate the broadening due to multiple scatterings in a more forward rapidity region, which is advocated in Ref.~\cite{Jia:2019qbl}. However, this may require accounting for saturation effects \cite{vanHameren:2019ysa} which together with the Sudakov effects act to generate considerable broadening in the final-state observables of the p--p and p--Pb collisions \cite{Aaboud:2019oop}. It will be interesting to see the combined effect of the two kinds of broadening.

\section*{Acknowledgements}
 KK, AvH and MR acknowledge the partial support by the National Centre of Science
(Narodowe Centrum Nauki) with the grant DEC-2017/27/B/ST2/01985. KT is supported by the Starting Grant from Trond Mohn Foundation (BFS2018REK01) and the University of Bergen.
\appendix

\section{Algorithm}
Here we outline the algorithm that simulates a hard-scattering process that yields two hard gluons which propagate as leading jet-particles through a medium. From a technical point of view, this algorithm merges two Monte Carlo programs: \katie\ and \mincas. 

First, \katie\ is executed.
The centre-of-mass energy per nucleon of the hard collision $\sqrt{s_{NN}}$ needs to be set ($2.76$~TeV for this work) as well as constraints for the minimum values of the $p_T$ of outgoing particles after the hard collisions. In general, the phase-space boundaries for the hard process, simulated by \katie, are set larger than those of the finally obtained set of particles (which involves \mincas\ as well). 
Also the factorisation scale is set and the sets of TMDs/PDFs are specified. As scale we use the average of transverse momenta of di-jets for this work.
  
Then, the following steps are repeated for each of the events stored in  the output files of \katie:  
\begin{enumerate}
\item From the \katie\ output files every event is read in individually. Essential for the algorithm are the outgoing-particle energy $E_i$ and three-momenta $\vec{q}_i$ (with $i=1,2$) as well as the weight $w$ of the event.
\item Polar coordinates for the outgoing-parton three-momenta $\vec{q}_i\equiv (q_{ix},q_{iy},q_{iz})$ ($i=1,2$) are calculated as:
\begin{align}
    q_i&=\sqrt{q_{ix}^2+q_{iy}^2+q_{iz}^2},\\
    \theta_i&=\arccos{q_{iz}/q_i},\\
    \phi_i&=\arctan{q_{iy}/q_{ix}}.
\end{align}
The outgoing particles in \katie\ are on the mass-shell, i.e.\ $E_i=q_i$.
\item The following steps are performed for each of the two outgoing particles $i=1,2$ individually:
\begin{enumerate}
\item Initialise \mincas\  for the particle $i$. There, as a single parameter from \katie, the particle energy $E_i$ (before propagation through the medium) is passed in order to calculate $t^\ast$ via Eq.~(\ref{eq:tstar}).
\item The generation of a \mincas-event -- the function {\tt MINCAS\_GenEve} is executed. 
\item {\tt MINCAS\_GenEve} yields the fraction $\tilde{x}_i$ with regard to the light-cone energy in the jet-frame\footnote{For this purpose, we define the jet-frame as the one obtained after the rotation of the coordinate system in the LAB frame, such that $\vec{q}_i$ is parallel to the $z$-axis of the new coordinate system.} $q_i^+= E_i$ 
as well as the momentum components $l_{ix}$ and $l_{iy}$ in the directions transverse to $\vec{q}_i$. Furthermore, to each jet the Monte Carlo weight $\omega_i$ is associated.
\item The energy $E_{p_i}$ and the momentum component $p_{iz}$ after the in-medium propagation are obtained in the jet-frame as:
\begin{align}
    E_{p_i}&=\tilde{x}_i q_i^+ + \frac{\ll_i^2}{4 \tilde{x}_i q_i^+}\,,\nonumber\\
    p_{iz}&=\tilde{x}_i q_i^+ - \frac{\ll_i^2}{4 \tilde{x}_i q_i^+}\,,
\end{align}
while the transverse components of $\vec{p}_i$ are given as $\ll_i=(l_{ix},l_{iy})$.
\item The new momentum $\vec{p}_i$ is rotated back into the LAB frame:
\begin{equation}
\left(
    \begin{array}{c}
        p_{ix}\\
         p_{iy}\\
         p_{iz}
    \end{array}
    \right)_{\rm lab}
    =
    \left(
    \begin{array}{ccc}
        \cos{\phi_i}&-\sin{\phi_i} &0  \\
         \sin{\phi_i}&\cos{\phi_i}&0\\
         0&0&1
    \end{array}
    \right)
    \left(
    \begin{array}{ccc}
    \cos{\theta_i}&0     &\sin{\theta_i}  \\
       0  & 1&0\\
       -\sin{\theta_i}&0     &\cos{\theta_i}
    \end{array}
    \right)   
     \left(
    \begin{array}{c}
         l_{ix}  \\
         l_{iy}\\
         p_{iz}
    \end{array}
    \right)_{\rm jet}.
\end{equation}
\item Then, $p_{iT}$ and $y_i$ in the LAB frame are obtained as:
\begin{align}
    p_{iT}&=\sqrt{p_{ix}^2+p_{iy}^2}\,,\\
    y_i&=\frac{1}{2}\log{\frac{E_{p_i}+p_{iz}}{E_{p_i}-p_{iz}} }\,.
\end{align}
\end{enumerate}
\item Finally, the event is written as the following three lines into the output-file:
\begin{align}
& E_1,\, q_{1x},\, q_{1y},\, q_{1z},\\
& E_2,\, q_{2x},\, q_{2y},\, q_{2z}, \\
& p_{1T},\,y_1, \,\phi_1, \,\tilde{x}_1,\,p_{2T}, \,y_2\,,\,\phi_2,\, \tilde{x}_2,\; \omega_1,\, \omega_2,\, w.
\end{align}
\end{enumerate}

\bibliography{refs}{}
\bibliographystyle{utphys_spires_tit}

\end{document}

%% file: definitions.tex
\usepackage{graphicx}
\usepackage{dcolumn}
\usepackage{bm}
\usepackage{epstopdf}
\usepackage{mathrsfs}
\usepackage{amsmath,amssymb,amsfonts,latexsym}
\usepackage{amsmath,bbold}
\usepackage{color}
\usepackage{slashed}
\usepackage{nicefrac}
\usepackage[colorlinks=true,linktocpage=true,linkcolor=blue,citecolor=blue]{hyperref}
\usepackage[utf8]{inputenc}

\def\bs{\boldsymbol}

\long\def\comment#1{ }

\def\0{{\boldsymbol 0}}
\def\q{{\boldsymbol q}}
\def\ll{{\boldsymbol l}}

\def\p{{\boldsymbol p}}

\newcommand{\beq}{\begin{eqnarray}}
\newcommand{\eeq}{\end{eqnarray}}
\newcommand{\bal}{\begin{align}}
\newcommand{\eal}{\end{align}}

\newcommand{\dd}{{\rm d}}

\newcommand{\nn}{\nonumber\\ }